\documentstyle[11pt,epsfig,here]{article}

\def\q2{$Q^2$}
\def\gev2{GeV$^2$}

\begin{document}
\vspace*{-1.8cm}
\begin{flushright}
\flushright{\bf LAL 99-69}\\
\vspace*{-0.5cm}
\flushright{December 1999}
\end{flushright}
\vskip 2 cm
\begin{center}
 {\LARGE\bf Inclusive Deep Inelastic Scattering at HERA \\
and related phenomenology } 
\end{center}

\vskip 1. cm

\begin{center}
{\bf\Large Fabian Zomer}
\end{center}

\begin{center}
{\bf\Large Laboratoire de l'Acc\'el\'erateur Lin\'eaire}\\
{IN2P3-CNRS et Universit\'e de Paris-Sud, BP 34, F-91898 Orsay Cedex}
\end{center}

\begin{abstract}
Recent measurements of inclusive deep inelastic
 scattering differential cross-section in the range
 $1.5$ \gev2$\le Q^2\le 30000$ \gev2 and $5\cdot 10^{-6}\le x\le 0.65$ 
are presented. Phenomenological analyses performed 
 from these measurements are also described. 
\end{abstract}

\section{Introduction}\label{introduction}

 In the Deep Inelastic Scattering (DIS) processes observed at HERA, a 
 lepton $\ell=e^\pm$ of \linebreak 27.5 GeV
interacts with a proton $P$ of 920 GeV yielding a lepton $\ell'$
 and a set of hadrons $X$ in the final state.
 Following the nature of $\ell'$ the interaction proceeds via a neutral
  ($\ell'=e^\pm$) current (NC) or a charged
 ($\ell'=\nu_e,\bar{\nu}_e$) current (CC). 
DIS events are collected in the H1 and ZEUS experiments \cite{nim-zeus-h1}
 which are located at the two $e^\pm P$ interaction points of HERA.  
 
 The kinematics of the DIS inclusive processes, 
$\ell(k)+ P(p) \rightarrow \ell'(k')+X$,  
 is determined by two independent kinematic variables, besides 
the energy of the incoming lepton and proton. One usually
chooses them among the four Lorentz invariants
\[
Q^2 \equiv -q^2=-(k-k')^2,\;
x=\frac{Q^2}{2p\cdot q}\,,\; y=\frac{p\cdot q}{p\cdot k},\;
 W^2=(q+p)^2,
\]
whereby at HERA one can neglect the lepton and proton masses so that 
 the useful relation $Q^2=xys$ holds. These kinematic 
 variables are obtained experimentally by measuring
 the momentum and/or the hadronic energy,
 the direction of the scattered lepton and/or the hadronic energy flow. 

 In this report we shall restrict ourselves to the cross-section
 measurements at HERA in the medium
 1.5 GeV$^2\le Q^2<150$ GeV$^2$ and high 150 GeV$^2\le Q^2\le 30000$ GeV$^2$ 
 domain of the DIS regime.
 During the past, a large number of precise measurements have been
 performed in the medium $Q^2$ region by 
fixed target experiments \cite{fixed-target-exp}.
 With HERA, three major improvements
 may be noticed:
\begin{itemize}
\item
 an extension of the \q2 domain
  to very high $Q^2$ ($10^4$ \gev2) but also to very
 small $x$ ($\approx 10^{-6}$);
\item
an almost hermetic ($4\pi$) detection of the final state
leading not only to the determination of the energy
 and angle of the scattered lepton but also of the produced hadrons; 
\item
from the previous items it follows that the detection
of both NC and CC is feasible in the same detector and during the same data
 taking period;
\end{itemize}

 The somewhat arbitrary distinction between low and high $Q^2$ 
 is related to different physics interests.
  In both regions perturbative Quantum-Chromo-Dynamics (pQCD) 
 is expected to describe the HERA data \cite{h1-zeus-94}. The pQCD analysis of 
medium \q2 data is part of a long tradition \cite{qcd-exp}
 from which 
the parton distributions of the nucleon and the strong
 coupling constant $\alpha_s$ have been extracted. On the top of that, very
 high \q2 ($\approx M_Z^2$) NC and CC   
data open a field of research in electroweak
 physics up to now reserved, in DIS, to neutrino beam experiments. 

 The rest of this report is organized as follows. In section 
 \ref{section_un} the measurements of NC and CC
 differential cross-sections are described. Section \ref{section_deux}
 is devoted to a phenomenological
 analysis of these measurements.

\section{Measurement of NC and CC cross-sections}
\label{section_un}

Neutral current events, at medium and high \q2,
 are basically identified by the presence of an
electron (or a positron) in the final state. This is done by  
using tracking and calorimetric devices covering the range 
 $7^o<\theta_e<177^o$ and $E'_e> $4 GeV(at HERA the forward
 direction $\theta_e=0^o$
 corresponds to the direction of the incoming proton).  

The differential cross-section measurement is done by counting the 
number of events within a kinematic interval in, say $x$ and 
\q2. Therefore one of the experimental problems is to achieve a good 
reconstruction of these kinematic variables from the detector information.
 Both, H1 and ZEUS,
 can use the outgoing lepton and hadronic final state information, namely the 
 polar angles, the momenta and the deposited energies. It is then possible
 to define the kinematics of each event by using different (and independent)
 combinations of experimental information. 

In ZEUS the double angle method is used 
\begin{eqnarray}
Q^2_{da}&=&4E_e^2\frac{\sin \gamma_h(1+\cos\theta_e)}
{\sin \gamma_h+\sin\theta_e-\sin(\gamma_h+\theta_e)},\,\,\,
y_{da}=\frac{\sin \theta_e(1-\cos\gamma_h)}{\sin \gamma_h+
\sin\theta_e-\sin(\gamma_h+\theta_e)}\nonumber\\
x_{da}&=&\frac{E_e}{E_p}\frac{\sin \gamma_h+
\sin\theta_e+\sin(\gamma_h+\theta_e)}{\sin \gamma_h+
\sin\theta_e-\sin(\gamma_h+\theta_e)}\nonumber .
\end{eqnarray}
 The hadronic polar angle $\gamma_h$ is defined by 
 $\tan\gamma_h/2=\sum_i(E_i-p_{z,i})/P_{t,h}$, where
 $E_i$ and $p_{z,i}$ 
 are the energy and longitudinal momentum of the final state hadron $i$ and where
 $P_{t,h}$ is the total transverse momentum of the hadronic final state particles.
In H1, the electron method is used
\[
Q^2_e=\frac{(E_e')^2\sin^2\theta_e}{1-y_e},\, \,\,
y_e=1-\frac{E'_e}{E_e}\sin^2(\theta_e/2)
\] 
to determine \q2 and $x$ only at $y>0.15$ since 
  $d x/x=1/y\dot d E'_e/E'_e$ while 
 at $y\le 0.15$ the $\Sigma$ method is used 
\[
Q^2_\Sigma=\frac{(E_e')^2\sin^2\theta_e}{1-y_\Sigma},\,\,\,
 y_\Sigma=\frac{\sum_i(E_i-p_{z,i})}{\sum_i(E_i-p_{z,i})+E'_e(1-\cos\theta_e)}\, .
\]
 The reason for the differences between the methods used by H1 and ZEUS are
 related to the calorimeter performances: H1 possesses finely segmented
 electro-magnetic calorimeters and ZEUS a very good hadronic calorimetry. 

 The redundancy in the determination of the kinematic variables is a
 crucial point and  presents
 many advantages: minimization of the migration between
 the `true' and the measured kinematic variable by choosing one particular
 method;
 cross calibration of the various calorimeter devices, and 
 studies of photon radiation from the
 lepton line by  comparing leptonic 
 and hadronic information.

Once the collected events are gathered in $x$-\q2 bins, besides the subtraction of
 photoproduction background, correction factors are applied for:
 the efficiency of the event selection; detector acceptance;
 wrong reconstruction of the kinematics due to detector effects, and the contribution
 of higher order electroweak processes. When possible, these correction
 factors are determined and/or cross checked from the data themselves. If this is not
 possible, then they are determined from a full simulation of the DIS and background
 processes including the detector response.
 
 For the medium \q2 data we shall describe the preliminary
 results of the high statistics
1997 data (more than 10000 events per $x$-\q2 bin for H1).  
For high \q2,  the combined 1994-1997\linebreak ($e^+$ beam) data published in
 ref. \cite{grand-Q2} are presented. 

At medium \q2 and for the H1 measurements,
 the main systematic uncertainties are: the electron energy scale ($\approx 0.3\%$),
 the hadronic energy scale ($\approx 2-3\%$), the electron polar angle\linebreak ($\approx 0.3 $ mrad), the photoproduction background at high-$y$  only ($\approx 3\%$ effect
 on the measurements) and 
 the correction factors
 (see above) applied to the data (each one is of the order of 1-2$\%$). 
 The overall data normalization (including the luminosity measurement) uncertainty
 is 1.5 $\%$.  The systematic uncertainty is, in total, of the order of 3$\%$ and is larger
than the statistical uncertainties which are at the level of 1 $\%$ for \q2$<100$ \gev2. 

At high \q2 the systematic uncertainties are essentially the same. In ZEUS 
 the statistic and systematic uncertainties amount to 3-5$\%$ for the kinematic
 range $400$ \gev2 $<Q^2<30000$ \gev2 considered in the analysis.

In charged current events,
 the outgoing neutrino escapes the detection. Such events  
 are then characterized by missing transverse energy $p_{t,miss}$.
Analyses in ZEUS (H1) have demanded $p_{t,miss}>10$ GeV ($p_{t,miss}>12$ GeV).
 For the reconstruction 
 of the kinematic variables, one can only use information
 from the hadronic final state, i.e. the Jacquet-Blondel method, giving, 
\[
y_{JB}=\frac{\sum_i(E_i-p_{z,i})}{2E_e},\,\,\, Q^2_{JB}=\frac{p_{t,miss}^2}{1-y_{JB}}
\, .
\]
The CC events statistics is still low,  $\approx$ 900 events for $Q^2>400$ \gev2 in ZEUS.
 The systematic uncertainty is dominated by the hadronic energy scale, which induces 
 an effect of the order of 10$\%$, except at very high \q2 
 and very high $x$ where the effect is above $20\%$. Other systematic sources related
 to the $p_{t,miss}$ cut, acceptance correction and photoproduction background subtraction 
(in the lowest \q2 bins) lead to measurement uncertainties between $4\%$ and $8\%$.

\section{Phenomenological analysis of inclusive measurements at HERA}
\label{section_deux}

As mentioned in the introduction, we shall distinguish the 
phenomenological analysis
of the medium \q2 data from the high \q2 data. As we are interested in the
 HERA data, it should be noted that we are considering the region
of large $W^2\gg 10$ \gev2. Therefore, we will not be concerned by the
 non-perturbative effects and the 
 higher twist effects appearing in this region so that the symbol pQCD,
 appearing below, refers to the leading twist of pQCD.

For all the mathematical details which cannot be given here we refer to ref. \cite{furmanski}
 and references therein.


\subsection{Analysis of the medium \q2 NC data}\label{qdc-medium}

In the one boson exchange approximation, the NC differential
 cross-section reads
\begin{equation}\label{nc-cross}
\frac{d\sigma^{e^\pm p}}{dxdQ^2}=\frac{2\pi\alpha_{em}Y_+}{xQ^4}\sigma_r,\,\,\,
\sigma_r=F_2(x,Q^2)-\frac{y^2}{Y_+}F_L(x,Q^2)\mp \frac{Y_-}{Y_+}xF_3(x,Q^2)\, ,
\end{equation}
 where $Y_\pm=1\pm (1-y)^2$. 
 The nucleon structure functions are modelled using the quark-parton 
 model and pQCD. In the so called naive parton model one writes 
\[
F_2(x)=\sum_{i=1}^{n_f}A_i(Q^2)x[q_i(x)+\bar{q}_i(x)]\,,\,
F_3(x)=\sum_{i=1}^{n_f}B_i(Q^2)[q_i(x)-\bar{q}_i(x)]  
\]
where $q_i$ ($\bar{q}_i$)is the density function of the quark
 (anti-quark) of flavor $i$,
 $n_f$ is the number of active flavors and 
$F_L=0$. The functions $A_i$ \cite{qcd-exp} depend 
 on the electric charge $e_i$ ($A_i=e^2_i$ for \q2$\ll M_Z^2$) and
 embody the effects of the $Z$ exchange and $\gamma-Z$ interference 
 in their \q2 dependence. The same holds for the functions $B_i$ \cite{qcd-exp} 
 except that they vanish at \q2$\ll M_Z^2$. 
 
 Going beyond the simple parton model, higher order contributions in $\alpha_s$
 are taken into account. However   
 mass singularities appear in the initial state of DIS processes and
 cannot be regularised without resuming the
 whole perturbative series. This resumation is done
 in a restricted 
 kinematic region where $\alpha_s\log Q^2$ is large \cite{furmanski}.
 This latter region is defined
 by\linebreak \q2$\gg\Lambda^2\approx 0.3^2$ GeV$^2$, and the pCDQ calculations
 are safe for \q2 above a few GeV$^2$.
 In this domain, the parton density functions 
(pdf) are given by the solution of the DGLAP equations \cite{furmanski}:

\begin{equation}\label{ap1}
\begin{array}{ll}
M_F\frac{\partial q^{\pm}_{i,NS}(x,M_F^2)}{\partial M_F}
&=P^\pm_{NS}\otimes q^{\pm}_{i,NS}(x,M_F^2)\nonumber \cr
M_F\frac{\partial}{\partial M_F }
\left (^
{\Sigma(x,M_F^2)}_{g(x,M_F^2)}\right )
&=
\left (^{P_{qq}  n_f P_{qg}}_
{P_{gq}   P_{gg}}\right )
\otimes
\left (^{\Sigma(x,M_F^2)}
_{g(x,M_F^2)}\right )
\end{array}
\end{equation}
 with $A\otimes B\equiv \int_x^1A(z)B(x/z){dz}/{z}$ and where  
 $\Sigma=\sum_{i=1}^{n_f} (q_i+\bar q_i)$
 is the singlet quark density, $ q^{-}_{i,NS}=q^v_i\equiv q_i-\bar q_i$
 and $ q^{+}_{i,NS}=q_i+\bar q_i-\Sigma/n_f$ are the two 
 non singlet densities and $g$ is the gluon density.
 The splitting functions $P_{i,j}=\alpha_s(M_R^2)P_{i,j}^{(0)}
+\alpha_s^2(M_R^2)P_{i,j}^{(1)}$ describe the branching of parton $j$
 from parton $i$, and they can be computed with pQCD up to the second order. 
  In eq. (\ref{ap1}) $M_F$ is the factorization scale
 (below which the mass singularity is resumed)
 and $M_R$ is the renormalisation
 scale (related to the ultra-violet singularity). 
 As the two scales must be chosen somehow,
 a natural choice for $M_F$ is $\sqrt{Q^2}$, i.e. the virtual mass
 of the probe. We shall, as usual, also set $M_R=M_F$ for convenience.
 It is worth mentioning that the DGLAP equations are universal, i.e. that they 
 are independent of the specific hard process.
 
 Eq. (\ref{ap1}) embodies the mass singularity resumation and therefore it 
 only describes the so called light parton, i.e. the parton of flavour $i$
 and mass $m_i$ such that $m_i^2/Q^2\ll 1$. In the medium \q2 range one can take
 the gluon, the up, down and strange quarks as the light partons. For the
 heavy quarks (charm and beauty) one needs to specify a special scheme. We 
 have chosen the fixed-flavor-scheme (FFS) \cite{ffs}
 -- suitable in the HERA medium \q2 range -- where beauty
 is neglected, and where the charm contribution is computed from the 
 boson-gluon-fusion process $\gamma g\rightarrow c\bar{c}$ plus the $\alpha_s^2$
 corrections. In this scheme charm is produced `outside' the hadron.
 The relation between the pdfs and the structure
 functions depends on the renormalisation
 scheme. In Next-to-Leading-Log-Approximation (NLLA)
 and in the $\overline{\rm{MS}}$ scheme one obtains: 
\begin{eqnarray}\label{f2_1}
 F_i(x,Q^2)=&x\sum_{j=1}^{n_f}
\biggl[
\biggl(
1+\frac{\alpha_s(Q^2)}{2\pi}C_{j,q}
\biggr)\otimes e_j^2(q_j(x,Q^2)+\bar{q}_j(x,Q^2))\nonumber\\  
&+2\frac{\alpha_s(Q^2)}{2\pi}C_{j,g}\otimes g 
\biggr]
+ F_i^{c\bar{c}}(x,Q^2)\nonumber
\end{eqnarray}
for $n_f=3$ and where $i=1,2$ (there is  
 a similar expression for $F_3$ with 
 $F_3^{c\bar c}=0$);
 $C_{j,q}$ and $C_{j,g}$ are the coefficient functions depending on the hard
 process; $F_i^{c\bar{c}}$ is the charm contribution 
 \cite{laenen}. It suffices here to say that it depends 
 on $m_c^2$ and on a renormalisation scale that we choose to be 
 $\sqrt{m_c^2+Q^2}$. Note that $F_L=F_2-2xF_1\ne 0$ in the NNLA. 

 To solve the system of integro-differential equations (\ref{ap1}),
 one must provide some initial conditions, i.e. some input functions of
 $x$ at a given \q2 for each pdf. 
 Since these
 functions reflect some unknown non-perturbative mechanism, one must
 parameterize with the help of a set of parameters. As we shall see below,
 these parameters are determined by comparing the calculations to
 the experimental data. 
 However, the 
 inclusive DIS data alone cannot constrain all light flavours since the
 structures functions are linear combinations of the pdfs:
introducing the singlet and a non-singlet
 $xu^+=xu+x\bar{u}-x\Sigma/3$ densities in order to write
 $\sum_{i=1}^{n_f=3}e_i^2x[q_i(x,Q^2)+\bar{q}_i(x,Q^2)]
={2}/{3}xu^++{1}/{9}x\Sigma$. 
 There is one important property of the DGLAP kernels $P_{i,j}$:
 the average total
 momentum carried by the partons, $\int_0^1(x\Sigma+xg)dx$, is independent of
 \q2. This quantity is called the momentum sum rule and is usually fixed to 1. 

 So in principle one may be able to describe the inclusive HERA data by
  parameterizing the three functions $xg$, $x\Sigma$ and $xu^+$.
 But we found that such a description
 is not adequate for the following reasons:
\vspace{2mm}
\begin{itemize}
\item
 it leads systematically to a too large fraction of the
 total momentum carried by the gluons $\int_0^1xgdx> 60\%$,
 in contradiction with the results of global fits including specific constrain 
 on $xg$ at high $x$ \cite{g_fit};
\item
 since the DGLAP equations involve some integrals of the pdfs from $x$ to 1,
 one must also introduce some constraints at higher $x$, i.e. the 
 fixed target hydrogen data from NMC and BCDMS \cite{fixed-target-exp};
\item
 even when fixed target hydrogen data are included, one is unable to constrain
 the total momentum carried by the gluons. One must in addition include the
 fixed target deuterium data. In this case a second non-singlet
 density must be parameterized (essentially $u+\bar u-d-\bar d$), but now 
 the valence counting rules, $\int_0^1(u-\bar u)dx=2$ and 
 $\int_0^1(d-\bar d)dx=1$, can be applied, under certain assumptions,
 in order to constrain the momentum sum rule at high $x$.
\end{itemize}
\vspace{5mm}
 H1 and ZEUS have used the latter option but with different assumptions.
 We will describe here the fits performed to the 1994 data and to the
 preliminary 1997 data. Both experiments include their own inclusive 
 measurements and the NMC and BCDMS data.

 In H1 two assumptions are made: $\bar u=\bar d$ and $\bar s=s=\bar d/2$.
 The first constraint is in contradiction with the global fit results 
 \cite{g_fit}
 including the Drell-Yann data but we have found that it does not modify 
 significantly the gluon density at below $x\approx 10^{-2}$. The 
second assumption comes from the results of the dimuon events of CCFR 
\cite{ccfr_dimuon}. Finally $xg$,
 $x\bar u$, $xu_v$ and $xd_v$ are parameterized at a given value
 of \q2 using the mathematical function $Ax^B(1-x)^CP(x)$ with
 $P(x)=1+Dx+E\sqrt{x}$. 
The momentum sum-rule and the quark counting rules are applied so that 
there is 16 free parameters in the H1 fit.
 
In ZEUS, the two valence quarks are taken from the MRS parameterization
 \cite{g_fit} and \linebreak $\bar s=s=(\bar d+\bar u)/2$ is also applied.  
 $xg$, $x(\bar u-\bar d)$ and $x(\bar u+\bar d)$ are parameterized
using the above mathematical functions.
  
Concerning the data-theory comparison, from which the input pdfs have to
be determined, both H1 and ZEUS use a $\chi^2$ minimization
procedure.
 The main steps of the fitting procedure are summarized below.
 For each iteration:
 1) the pdf's are parameterized at a given value of \q2 
 denoted $Q^2_0$, it is chosen to be 7 \gev2 in the ZEUS fit and
 4 \gev2 in the H1 fit, 
 2) the DGLAP equations are solved numerically in the $x$-space 
\cite{method}.
3) the evolved pdf's are convoluted with the 
 coefficient functions to obtain the structure functions. 
 Assuming that all experimental
uncertainties are normally distributed a $\chi^2$ is 
computed. A crucial point of the analysis is the $\chi^2$
expression which permits the use of the correlations introduced
 by some of the systematic uncertainties:
 \begin{eqnarray}\label{lechi2}
\chi^2
&=&\sum_{exp} \sum_{dat}
 \frac{[{\cal O}^{dat}_{exp}-
{\cal O}^{ fit}\times(1-\nu_{exp}\sigma_{exp}
-\sum_k\delta^{dat}_k(s^{exp}_k))
]^2}
{\sigma_{dat,stat}^2+\sigma_{dat,uncor}^2}\nonumber\\
&+&\sum_{exp}\nu_{exp}^2  
+\sum_{exp}\sum_k (s^{exp}_k)^2 \nonumber 
\end{eqnarray}
where ${\cal O}$ stands for the observables  
(structure functions and differential cross-sections). 
The first two sums run over the data ($dat$) of the
various experiments ($exp$); 
       $\sigma_{exp}$ is the relative overall normalization uncertainty
 of the experiment $exp$; 
 $\sigma_{dat,stat}$ and $\sigma_{dat,uncor}$ are the statistical error 
 and the uncorrelated systematic error, 
respectively,  corresponding to the data point $dat$;
 $\nu_{exp}$ is the number of standard deviations corresponding to
 the overall normalization of the experimental sample $exp$;
 $\delta^{dat}_k(s^{exp}_k)$ is the relative shift of the data point $dat$
 induced by a change by $s^{exp}_k$ standard deviations of
 the $k^{th}$ correlated systematic uncertainty source
  of the experiment $exp$. It is estimated by 
\begin{eqnarray}
\delta^{dat}_k(s^{exp}_k)=
\frac{{\cal O}^{dat}_{exp}(s^{exp}_k=+1)-{\cal O}^{dat}_{exp}(s^{exp}_k=-1)}
{2{\cal O}^{dat}_{exp}}\, s^{exp}_k+\nonumber\\
\biggl[
\frac{{\cal O}^{dat}_{exp}(s^{exp}_k=+1)+{\cal O}^{dat}_{exp}(s^{exp}_k=-1)}
{2{\cal O}^{dat}_{exp}}-1
\biggr] \,(s^{exp}_k)^2\,,\nonumber
\end{eqnarray}
where ${\cal O}^{dat}_{exp}(s^{exp}_k=\pm1)$ is the experimental determination
of ${\cal O}^{dat}_{exp}$ obtained varying by $\pm 1 \sigma$ the $k^{th}$
source of uncertainty.
  Parameters $\nu_{exp}$ and $s^{exp}_k$ can be determined 
 by the
 $\chi^2$ minimization or they can be fixed to zero during the 
 minimization but released during the $\chi^2$ error matrix calculation.
 In the first case one uses all the experimental information relying on 
 the correctness of the estimate of the systematic uncertainties.

 The result of the H1 fit is shown
 in fig. \ref{f2-fig} together with the data. The 
 agreement between data and pQCD is excellent.
 The gluon density obtained from the 
 ZEUS fit (to the 1994 data) is shown in fig. \ref{gluon-fig}.
 The error bands of the gluon density  
 include the experimental error propagation
 as defined in ref. \cite{nous} and a theoretical uncertainty
 which includes the variation of: $\alpha_s$,
 the charm mass, 
 the pdf input parameterization form, the
 value of $Q_0^2$ and the factorization scale. With the 1997 data
 one can expect
 a reduction of the experimental uncertainty by a factor of two.  
 The theoretical
 uncertainties will then dominate in the determination of $xg$, i.e.
 the third order splitting functions are needed.  
\begin{figure}[h!]
\begin{center}
\includegraphics[width=.7\textwidth]{zomer.1}
\end{center}
\vspace{-0.5cm}
\caption[]{H1 preliminary measurements of $\sigma_r$ together with 
 the result of a pQCD fit (see text).}
\label{f2-fig}
\vspace{-0.2cm}
\begin{center}
\includegraphics[width=.4\textwidth]{zomer.2}
\end{center}
\vspace{-0.5cm}
\caption[]{$xg$ extracted from the ZEUS fit to the 1994 data for three
 values of \q2. Error bands contain the experimental error propagation and 
 a theoretical error estimation (see text).}
\label{gluon-fig}
\end{figure}

\begin{figure}[h]
\begin{center}
\includegraphics[width=.8\textwidth]{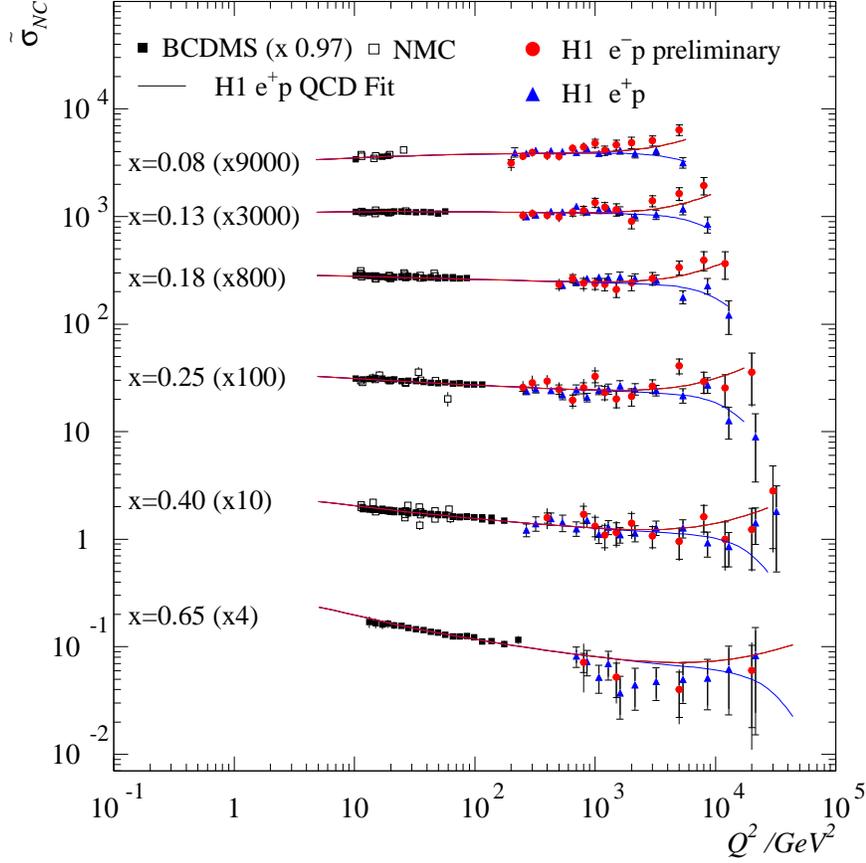}
\end{center}
\caption[]{High \q2 H1 measurements of $\tilde\sigma\equiv\sigma_r$
 (eq. \ref{nc-cross}) compared with 
 pQCD fit results (see text).}
\label{nc-fig}
\end{figure}
\begin{figure}[h]
\begin{center}
\includegraphics[width=.6\textwidth]{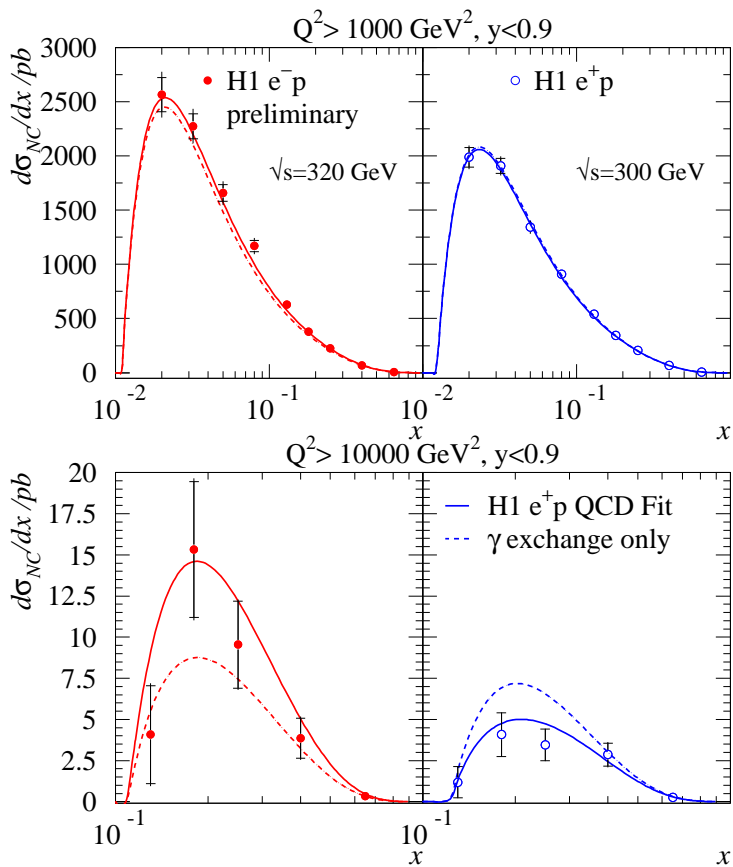}
\end{center}
\caption[]{H1 measurements of $d\sigma^{NC}/dx$ together with 
 results from pQCD fits and different assumptions on electroweak contributions
 (see text).}
\label{mz-effects}
\end{figure}
\begin{figure}[h!]
\begin{center}
\includegraphics[width=.6\textwidth]{zomer.5}
\end{center}
\vspace{-5mm}
\caption[]{ZEUS measurements of $\tilde\sigma\equiv\Phi_+$ together with 
 various pQCD calculations (see text).}
\label{cc-fig}
\begin{center}
\includegraphics[width=.6\textwidth]{zomer.6}
\end{center}
\vspace{-10mm}
\caption[]{H1 measurements of $d\sigma^{CC}/dQ^2$ together with the
 pQCD fit results.}
\label{q2-slope-fig}
\end{figure}
\begin{figure}[h]
\begin{center}
\includegraphics[width=.5\textwidth]{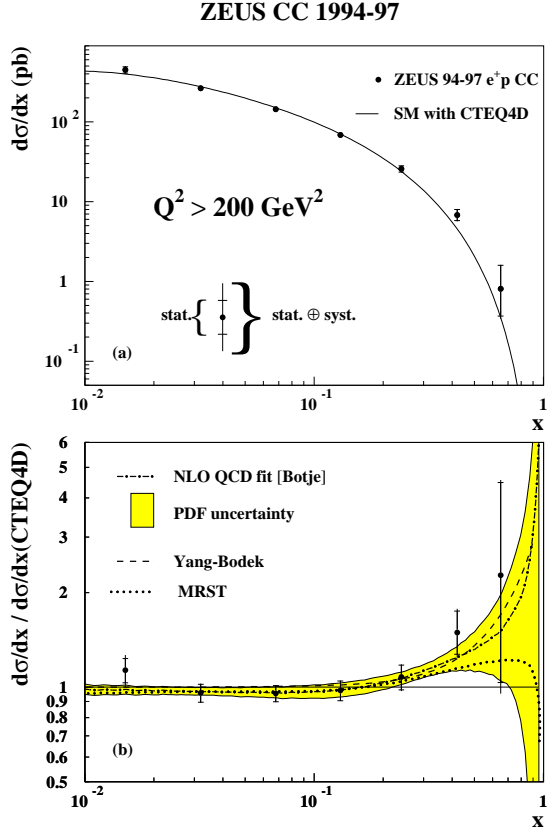}
\end{center}
\vspace{-10mm}
\caption[]{ZEUS measurements of $d\sigma^{CC}/dx$ (for $e^+p$) together with 
 various pQCD results (see text).}
\label{ccx-fig}
\end{figure}
\subsection{Analysis of the high \q2 NC and CC data}\label{qdc-high}

 The fits applied to the high \q2 data differ from the one described 
 in the previous section by a different calculation of the 
  contribution of the heavy 
 quarks to the structure functions. As $m_c\approx 1.5$ GeV,
 one has $m_c/Q^2\ll 1$ at high \q2. The large term
$\alpha_s^n\log^n(Q^2/m_c^2)$ 
 -- dominating the calculation of $F_2^{c\bar c}$ --
 must be resumed already at \q2$\approx 20$ \gev2. 
 The massless scheme is therefore used and only data with
 $Q^2\ge10$ \gev2 are included in the fit.
 In the massless scheme, charm and beauty are considered as partonic 
 constituents of the proton
 and their density functions are obtained by solving the DGLAP equations 
 with the initial conditions $c(x,Q^2\le m^2_c)=0$ and $b(x,Q^2\le m^2_b)=0$.  
 Such  fits describe the HERA NC and CC
 (see fig. \ref{nc-fig} and \ref{cc-fig}) data above \linebreak \q2= 10 \gev2. 

 In fig. \ref{nc-fig} one can observe the different behaviour of $e^-p$ 
 and $e^+p$ cross-sections at very high \q2. This is related to the different
 sign of the contributions of $F_3$ to $\sigma_r$. 
 Fig. \ref{mz-effects} shows $d\sigma/dQ^2$ together with the results of two
 pQCD fits including or not the $Z$ exchange and $\gamma -Z$ interference.
 With the present data, sensitivity to electroweak effects in NC
 is for the first time observed at HERA.

\vspace{10mm}
Up to now we have only described the NC cross-sections and related
 structure functions. For CC processes, in the one boson exchange
 approximation, one has
\begin{equation}\label{cross-cc}
\frac{d\sigma_{CC}^{e^\pm p}}{dx dQ^2}=\frac{G_F^2}{2\pi x}
\frac{M_W^4}{(M_W^2+Q^2)^2}\Phi_\pm(x,Q^2),
\end{equation}
where $G_F$ is the Fermi constant, and where the functions $\Phi_\pm$ 
depend on CC structure functions  (see \cite{qcd-exp} for example).
From eq. (\ref{cross-cc}) one can first remark  that the \q2 slope of the CC differential cross-section
 (see fig. \ref{q2-slope-fig})
permits a determination of $M_W$, assuming (or not) the precisely measured value
 for $G_F$ \cite{pdg}.
 To extract  $M_W$, H1 and ZEUS have used two different procedures. In H1, $M_W$
 is taken as an extra free parameter ($G_F$ is fixed) of the pQCD fit 
and in ZEUS,
 the pdfs of CTEQ \cite{g_fit}
 are used in order to extract $M_W$ and $G_F$ (variations of the pdf choice
 is taken into account within the errors). The results are
\begin{eqnarray}
\rm{H1:}\,\,
  M_W&=&80.9\pm 3.3(stat.)\pm 1.7 (syst.)\pm 3.7 (theo)\, \rm{GeV}\nonumber\\
\rm{ZEUS:}\,\, 
 M_W&=&80.4^{+4.9}_{-2.6}(stat.)^{+2.7}_{-2.0}(syst.)^{+3.3}_{-3.0}(pdf)\, \rm{GeV}\nonumber
\end{eqnarray}
\newpage
and treating $G_F$ as free, ZEUS obtains:
\begin{eqnarray}
  M_W&=&80.8^{+4.9}_{-4.5}(stat.)^{+5.0}_{-4.3}(syst.)^{+1.4}_{-1.3}(pdf)
 \,\rm{GeV},
\nonumber\\ 
 G_F&=&[1.171\pm0.034(stat.)^{+0.026}_{-0.032}(syst.)^{+0.016}_{-0.015}(pdf)]
\times 5\cdot 10^{-5}\,\rm{GeV}^{-2}.\nonumber
\end{eqnarray}

 Let us point out that, concerning the H1 result, the theoretical uncertainty
is dominated by the variation of the results when varying the ratio
 $\bar d/\bar u$ in the pQCD fit, and by the choice of
  the nuclear 
corrections applied to the deuterium target data entering the fit.
 These results,
 in good agreement with the world average values \cite{pdg}, show that
 the standard model gives a good description of both space-like (CC in DIS)
 and time-like ($W$ production in $p\bar p$ and $e^+e^-$ collisions)
 processes.

 In order to see the sensitivity of the CC cross-section to the pdfs,
 we write $\Phi_\pm$ in LO
\begin{equation}
\Phi_+=x\bar U+(1-y)xD;\,\,\, \Phi_-=xU+(1-y)x\bar D\nonumber
\end{equation}
with $U=u+c$ and $D=d+s$. From these expressions and from fig. \ref{cc-fig}
 one can remark that:
 with positron (electron) beams one can determine $d^v$ ($u^v$) at high $x$ and
 small-$y$ and
 $\bar u +\bar c$ ($\bar d+\bar s$) at small $y$. Let us mention that $d_v$
 and the sea quarks 
 are basically determined in the global pQCD fits
 by $\mu d$ and $\nu(\bar \nu)F_e$ 
 fixed target data, which require some nuclear corrections. Therefore, with 
 the HERA $e^{\pm} p$ CC events one may have, with more statistics, a unique 
 means to determine properly these quark densities.
\newpage
\vspace*{-0.5cm}
 In fig \ref{ccx-fig},
 $d\sigma/dx$ is 
shown together with the error band determined by the ZEUS pQCD fit
 (without the CC and NC
 data described in the present article), and with the results of a
 recent analysis where
 an ansatz $d/u\ne 0 $ as $x\rightarrow1$ \cite{bodek} was introduced. 
 Although the statistics is still low, one can notice 
from fig. \ref{ccx-fig}
 that this latter hypothesis is not required by the HERA data.
In fig. \ref{new-cc}, the preliminary 1998 measurement of 
$d\sigma^{e^-p}/dx$  is shown.
 The error band of the pQCD is much smaller than in fig. \ref{ccx-fig},
 therefore one can expect a better determination of electroweak parameters.
 The size of the error bands reflect that
 $u^v$ is much better constrained than $d^v$ in the pQCD fits.

\subsection{Extraction of $F_L$}\label{section-fl}
The longitudinal structure function is very hard to determine. It requires to 
 combine data in a given $x$-\q2 bin from different beam energies. However,
 from eq. (\ref{nc-cross}), one observes that at high $y$ the cross-section
 receives 
a contribution both from $F_2$ and $F_L$. Therefore, taking $F_2$ from the 
result of a pQCD fit (see previous section)
 applied to the low $y$ ($y<0.35$) data one can determine
 $F_L$ at high $y$ 
 by subtracting $F_2$, extrapolated to high $y$.
 The result of this operation is shown in fig. \ref{fl-fig}. To reach lower 
\q2, where pQCD is not reliable, another method is used.  Writing 
\[
\frac{\partial \sigma_r}{\partial \log y}=\frac{\partial F_2}{\partial \log y}
-2y^2\frac{2-y}{Y_+^2}F_L-
\frac{y^2}{Y_+}\frac{\partial F_L}{\partial \log y},
\]
 neglecting $\partial F_L/\partial \log y$, and assuming that
 $\partial F_2/\partial \log y$
 is a linear function of $\log y$, one obtains the results also shown in fig.
 (\ref{fl-fig}).
 This determination is consistent with the LO calculation of pQCD.
\newpage
\begin{figure}[h!]
\begin{center}
\includegraphics[width=.6\textwidth]{zomer.8}
\end{center}
\vspace{-10mm}
\caption[]{ZEUS measurements of  $d\sigma^{CC}/dx$ (for $e^-p$)
 together with the pQCD fit result.}
\label{new-cc}

\vspace*{2mm}
\begin{center}
\includegraphics[width=.6\textwidth]{zomer.9}
\end{center}
\vspace{-5mm}
\caption[]{H1 determination of $F_L$.
 The full points correspond to
 the subtraction method and the stars to the derivative method (see text).}
\label{fl-fig}
\end{figure}

\section{Conclusion}

Recent measurements of medium and high \q2 differential cross-sections  
 at HERA have been presented, and a 
 determination of $F_L$ was also described.
 These new inclusive DIS data cover four orders of magnitude in \q2
 and five orders of magnitude in $x$.  

In order to test Quantum Chromodynamics,
 fits based on the DGLAP equations have
 been performed successfully to the NC and CC data sets
 presented in this article. The extraction of the gluon density was described
 and the result of the analysis of the 1994 data was shown.

 At medium \q2 HERA has reached the limit where the
 systematic uncertainties dominate the statistical ones. 
 With the 1997 data, the gluon density 
 will be determined by the pQCD fit at the 
few percent level of accuracy, and a 
 determination of both $xg$ and $\alpha_s$, is now foreseen.

 Concerning the high \q2 data, although the statistics of NC and CC events
 is still low, a sensitivity of the results to the effects of $Z$ boson
 and $\gamma-Z$ interference in NC and to $M_W$ in CC was observed.
 A determination of $M_W$ from  
 $d\sigma^{CC}/dQ^2$ in space-like DIS was reported and a good agreement
 was found with the world average value from measurements in the time-like
 regions. Furthermore, comparing
 $d\sigma^{CC}/dxdQ^2$ with the pQCD calculation, we pointed out that
 such a measurement offers a unique possibility to pin down --
 independently of any nuclear effects --
 $d^v$ and the different components of the proton sea.

 Finally, comparing the measurements of $d\sigma^{e^+p}/dxdQ^2$ and
 $d\sigma^{e^-p}/dQ^2$ we observed, for the first time, the sensitivity
 of the NC to $F_3$.

 With the high \q2 NC and CC events a new field of research is touched here.
 It will be covered, with more precision,
 by the HERA-2000 upgrade with the help
 of an increase of luminosity and longitudinal
 polarization of the lepton beam.

%


\begin{thebibliography}{8.}
\addcontentsline{toc}{section}{References}

\bibitem{nim-zeus-h1}H1 Coll., Nucl. Instr. Meth A336, \textbf{310}
 and \textbf{348} (1997); The ZEUS detector, status report DESY-1993.
\bibitem{fixed-target-exp} BCDMS Coll., Phys. Lett. 
 B223, \textbf{485} (1989); Phys. Lett. B237, \textbf{592} (1989); 
 NMC Coll., Nucl. Phys. B\textbf{483}, \textbf{3} (1997).
\bibitem{h1-zeus-94} H1 Coll. Nucl. Phys. B470, \textbf{3} (1996);
  ZEUS Coll. Eur. Phys. J. C7, \textbf{609} (1999),
 see also \linebreak M. Botje, DESY 99-038 for the ZEUS pQCD fits.
\bibitem{qcd-exp} See for a recent review:
 A.M. Cooper-Sarkar, R.C. Devenish and A. De Roeck, 
  Int. J. Mod. Phys. A13, \textbf{3385} (1998).
\bibitem{grand-Q2} H1 Coll., DESY 99-107; ZEUS Coll., DESY 99-56 and 
 DESY 99-59. 
\bibitem{furmanski} W.~Furmanski and R.~Petronzio, 
Z. Phys. C1, \textbf{293} (1982).
\bibitem{ffs} M.~Gluck, E.~Reya and M.~Stratmann, 
 Nucl. Phys. B422, \textbf{37} (1994).
\bibitem{laenen}
E.~Laenen et al., Nucl. Phys. B392, \textbf{162} and \textbf{229} (1993);
 Phys. Lett. B291, \textbf{325} (1992).
 \bibitem{g_fit} CTEQ4 Coll., Phys. Rev. D55, \textbf{1280} (1997);
 MRS Coll., Phys. Rev. D51, \textbf{4756} (1995).
\bibitem{ccfr_dimuon}  CCFR Coll., 
 Z. Phys. C65, \textbf{189} (1995).
\bibitem{method}  C.~Pascaud and F.~Zomer, DESY 96-266; 
 J. Bl{\"u}mlein et al., in 
Proc. of the Workshop on Future Physics at HERA, 
  G.~Ingelman, A.~De Roeck and R.~Klanner eds., \textbf{23} DESY (1996).
\bibitem{nous}  C. Pascaud and F. Zomer, LAL 95-05.
\bibitem{pdg} Particle Data Group, Eur. Phys. J. C3, \textbf{1} (1998).
\bibitem{bodek} U.K. Yand and A. Bodek, Phys. Rev. Lett. 82, \textbf{2467}
 (1999).
\bibitem{hera-2000} See for instance
 Proc. of the Workshop on Future Physics at HERA, 
  G.~Ingelman, \linebreak A.~De Roeck and R.~Klanner eds., DESY (1996).

\end{thebibliography}
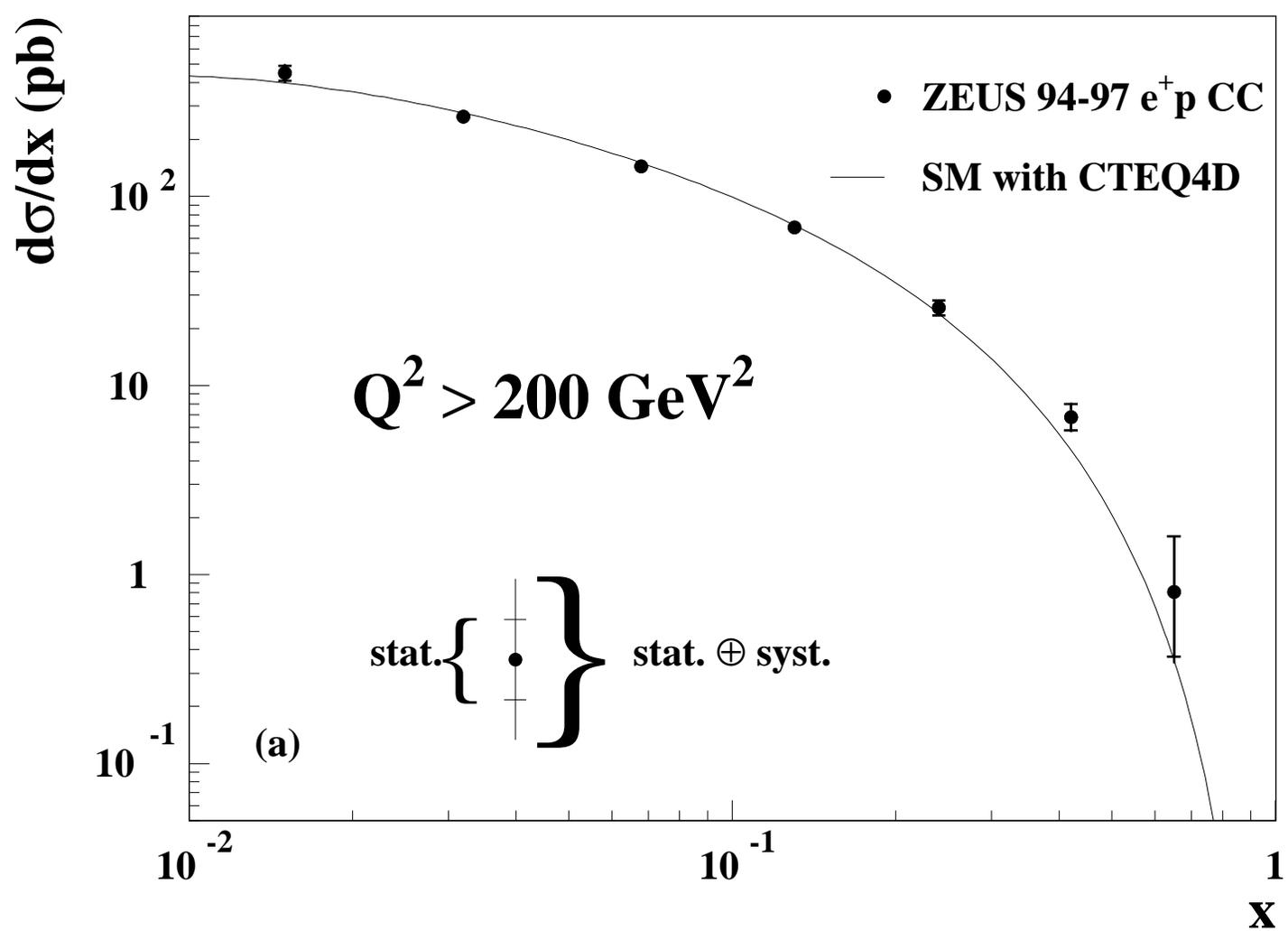
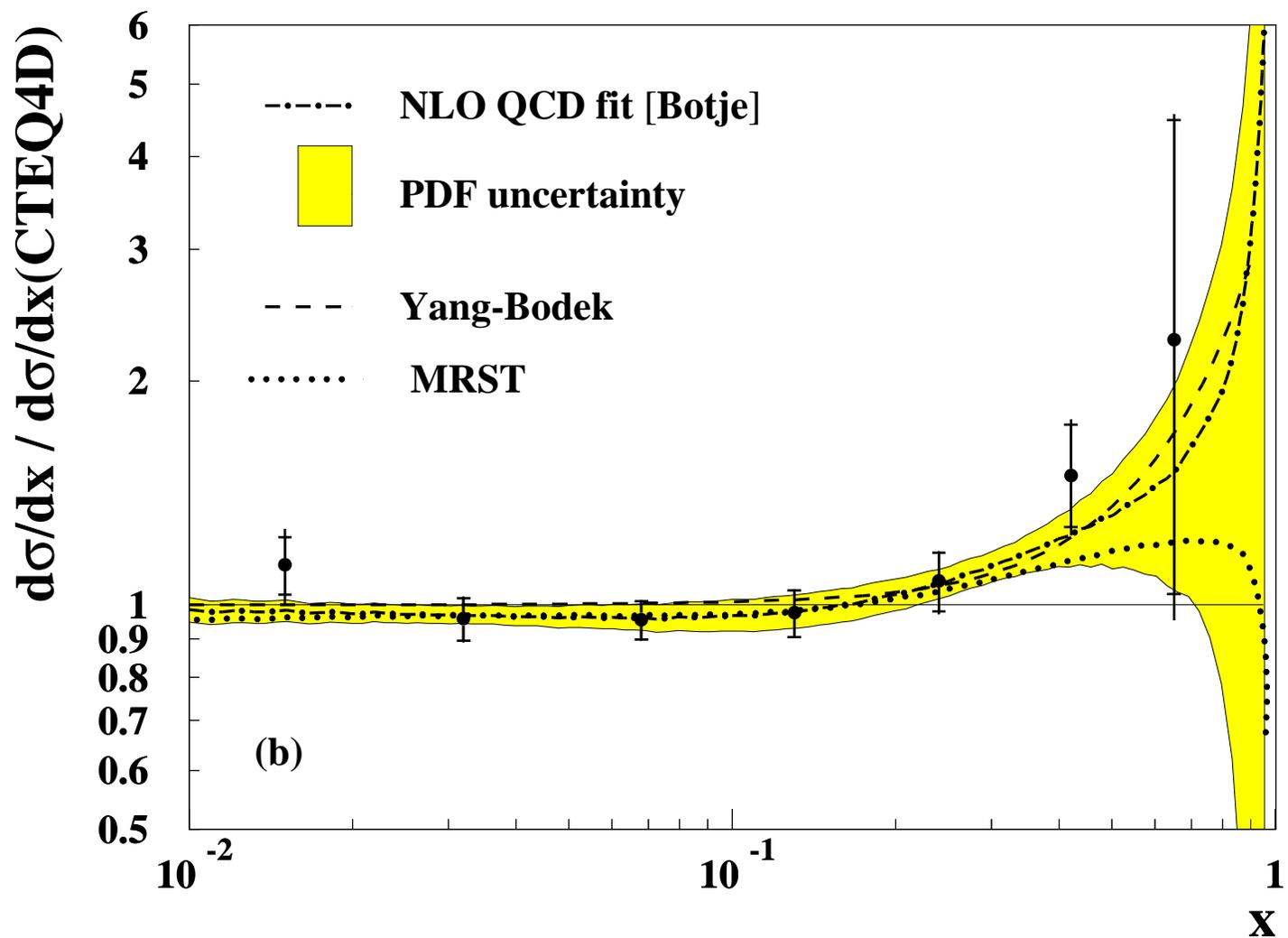
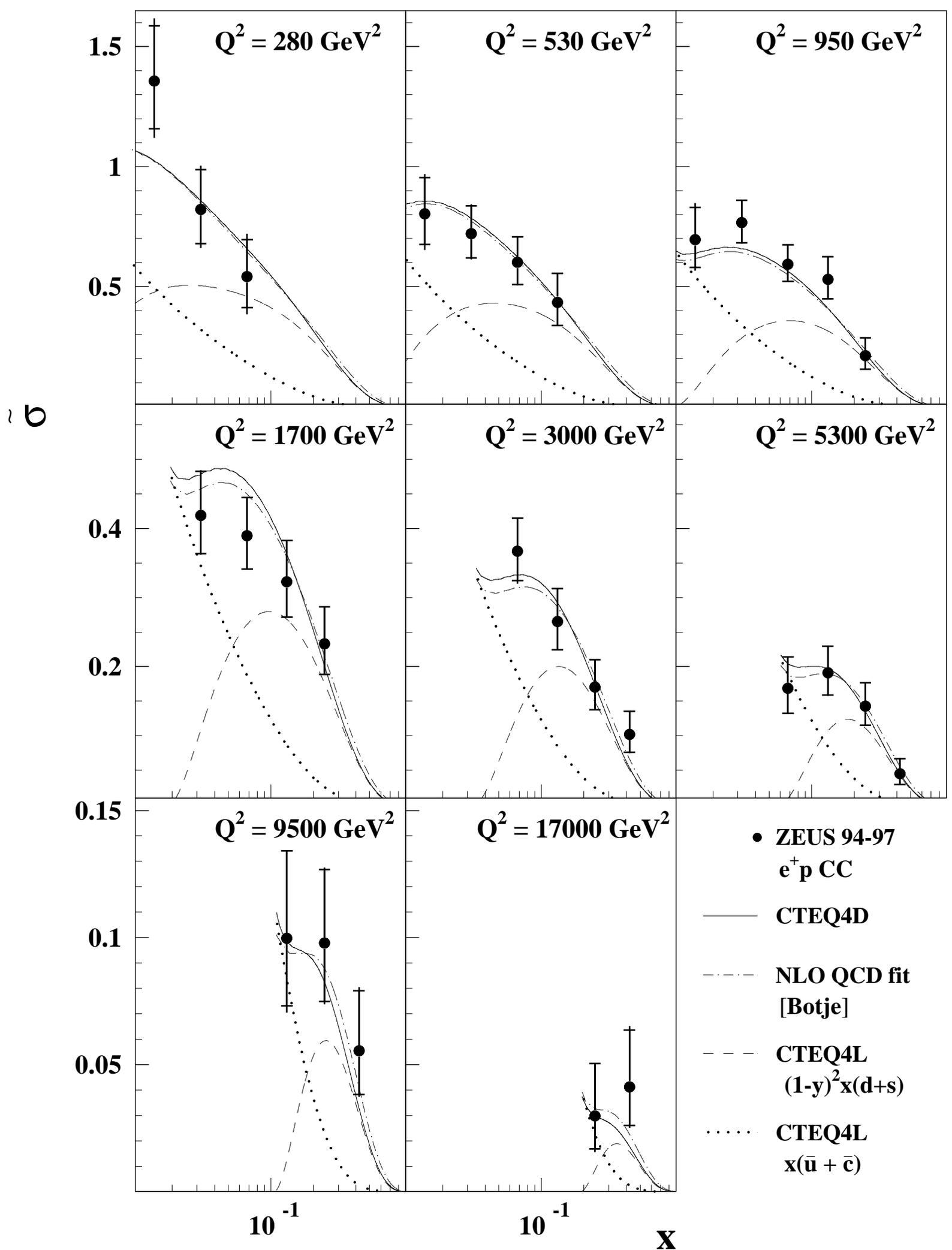
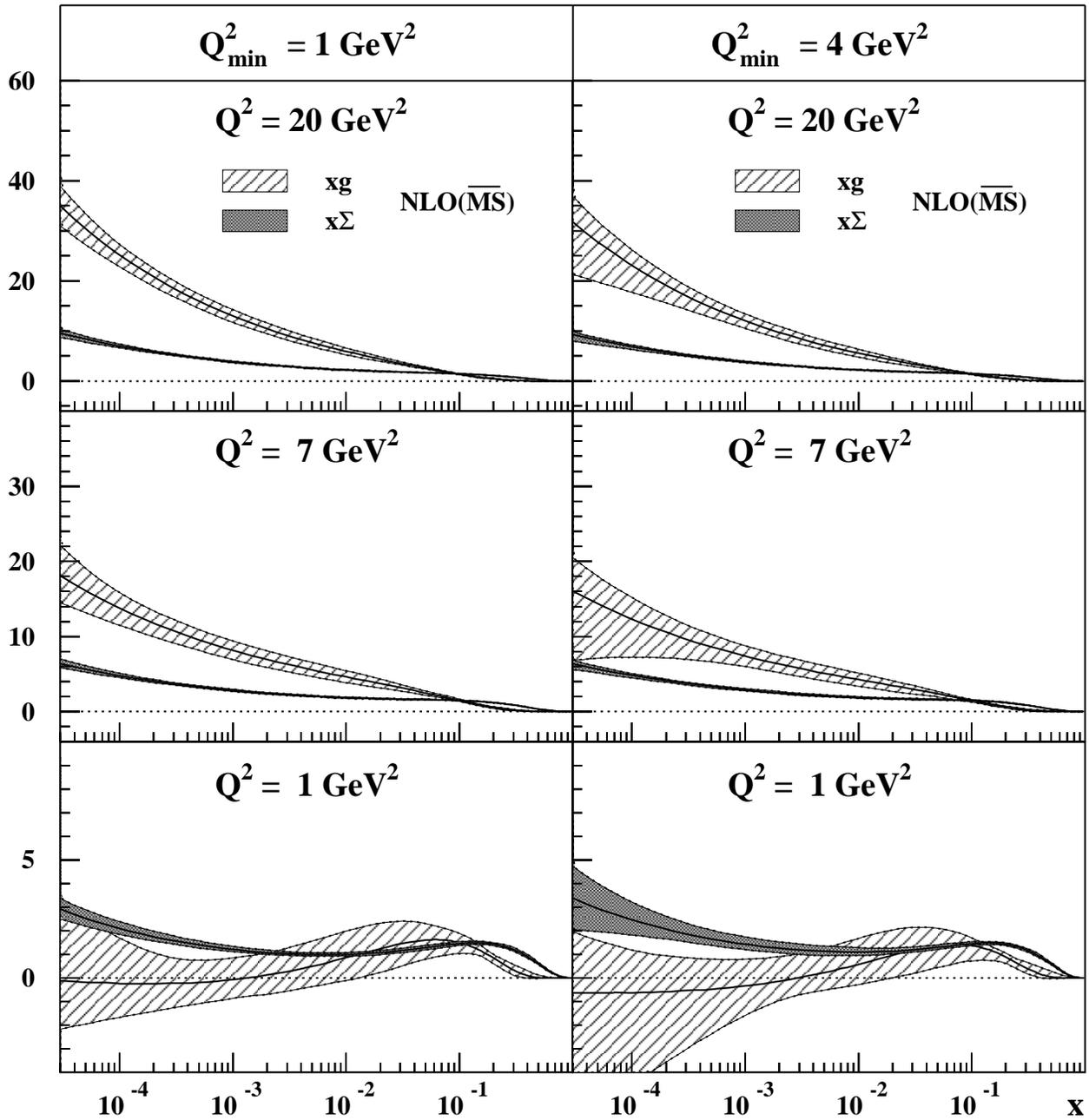
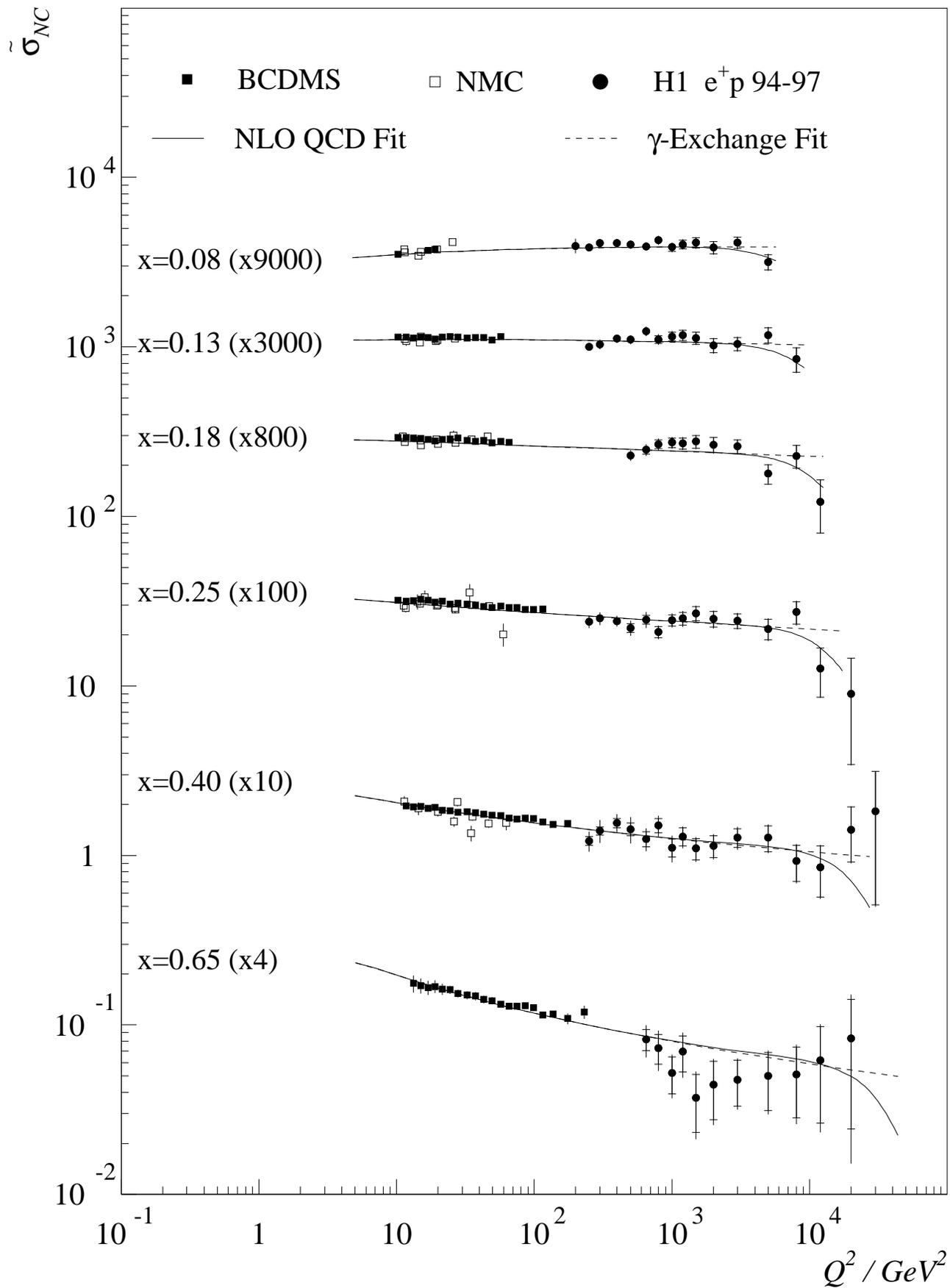
\end{document}